\documentclass[14pt]{article}
\usepackage{arxiv}
\usepackage{amsmath,amssymb}
\usepackage{caption}
\usepackage[utf8]{inputenc} 
\usepackage[T1]{fontenc}    
\usepackage{hyperref}       
\usepackage{url}            
\usepackage{booktabs}       
\usepackage{amsfonts}       
\usepackage{nicefrac}       
\usepackage{microtype}      
\usepackage{lipsum}
\usepackage{placeins}
\usepackage{graphicx}
\usepackage{multirow}
\usepackage{subcaption}
\usepackage[
    backend=biber,
    style=authoryear,
    doi=true
]{biblatex}
\graphicspath{ {./images/} }
\numberwithin{equation}{section}
\usepackage{setspace}
\usepackage{lineno}

\title{\texttt{poscosea} : A Computationally Efficient Sensitivity Analysis for Bayesian Models using the posterior covariance representation}

\author{
 Yusaku Ohkubo \\
  Okayama University\\
  Okayama, Japan \\
  \texttt{y-ohkubo@okayama-u.ac.jp} \\
         \And
   Yukito Iba \\
  The Institute of Statistical Mathematics\\
  10-3 Midori cho, Tachikawa City, Tokyo 190-8562, Japan \\
  \texttt{iba@ism.ac.jp} \\
}

\begin{document}
\maketitle

\section*{Acknowledgements}
This research was partially supported by JSPS KAKENHI (Grant Numbers 21K15170 and 24K15120). The authors appreciate Ph.D. Natsuki Ogusu (RIKEN) and Associate Prof. Ph.D. Keisuke Yano (the Institute of Statistical Mathematics) for providing valuable feedback. We partly used ChatGPT for language editing.

\section*{Data Availability}
Empirical data for Real Data Analysis section is available from the R package \texttt{AHMbook}.

\section*{Conflicts of Interest}
The authors declare no conflicts of interest.

\section*{Author Contribution}
YI conceived the ideas and designed methodology. YO conducted numerical simulations, data analysis, implementation of the software, and led the writing of the manuscript. All authors discussed the results, contributed to the drafts and gave final approval for publication.
\clearpage

\begin{abstract}
\begin{enumerate}
  \item Bayesian methods are essential in modern data analysis in ecology and evolutionary biology. They provide a flexible framework for modeling complex data-generating processes, while quantifying uncertainty based on the classical subjective interpretation of probability. 
  \item However, Bayesian inference may provide misleading measures of uncertainty, particularly when the fitted model fails to adequately represent the true data-generating process. Although nonparametric approaches such as leave-k-out diagnostics and bootstrap resampling offer more robust alternatives under model misspecification, their computational cost is often too demanding because they require repeatedly refitting the same Bayesian model. 
  \item In this paper, we introduce posterior covariance sensitivity analysis (PosCoSeA), a computationally efficient strategy for approximating leave-k-out diagnostics and bootstrap resampling without repeated model refitting. The performance of the methods is evaluated through both simulation studies and an application to real ecological data.
  \item We also provide an \textsf{R} package that implements these methods to facilitate their practical application. 
\end{enumerate}
\end{abstract}


\fontsize{12}{13.5}\selectfont

\section{Introduction}
Bayesian inference has gained widespread popularity in ecology over the past decades (\cite{mccarthy2007bayesian}; \cite{kery2011bayesian}; \cite{kery2020applied}). The development of user-friendly software implementing Markov chain Monte Carlo (MCMC) algorithms, such as WinBUGS, JAGS, and Stan, has facilitated the application of Bayesian methods in ecological studies. Consequently, a wide variety of Bayesian models have been developed, including occupancy models for species distribution (\cite{royle2007bayesian}), capture–mark–recapture models for survival estimation (\cite{dupuis1995bayesian}), state-space models for population dynamics (\cite{de2002fitting}), and mixed models incorporating structured random effects to account for spatial autocorrelation, temporal dependence, and phylogenetic relatedness (\cite{hadfield2010general}, \cite{ohkubo2023novel}). These models frequently employ hierarchical parameter structures, enabling researchers to account for multiple sources of uncertainty and variation while representing complex ecological processes. While Bayesian statistics are often founded on the degree-of-belief theories of probability (e.g., de Finetti, Savage), which gives an interpretation of probability to represent uncertainty about a result, many practitioners are embracing the more “pragmatic” advantages of flexible Bayesian modeling (\cite{gelman2013philosophy}), sometimes using a hybrid strategy that provides Frequentist justification for Bayesian methods (i.e., evaluating the expected behavior of a Bayesian method under repeated sampling from a data-generating process; \cite{mayo2018statistical}, \cite{ohkubo2021revisiting}).

Despite its popularity, however, Bayesian inference poses difficulties in several situations because it can yield biased point estimates or its uncertainty representation (e.g. posterior standard deviation or credible interval) can be misleading (\cite{lele2009bayesian}). For example, previous studies have revealed that N-mixture, one of the most popular hierarchal model to estimate a population abundance with imperfect detection process yields biased estimate of unknown parameters when the replication of survey per site or the number of site is small (\cite{knape2015estimates}, \cite{yamaura2016study}, \cite{kery2020applied}) due to the weak identifiability. In these cases, posterior inference can become highly sensitive to prior assumptions and to a small number of influential observations. The exclusion of even a single influential observation may alter posterior estimates and associated credible intervals. The situation will be further complicated when the model is misspecified (\cite{knape2018sensitivity}); that is, when the assumed statistical model does not adequately represent the true data-generating process. For example, models may be misspecified when the true relationship between predictors and the response is nonlinear, when observations exhibit unmodeled overdispersion, or when important covariates are omitted from the model. Under model misspecification, posterior distributions may yield biased point estimates and credible intervals with poor Frequentist coverage. Assessing reliability of the results is fundamental to offer scientific evaluation of a hypothesis, prediction for practitioners, and evidence for decision makers.

Several strategies have been proposed to identify influential observation and obtain a more reliable assessment of uncertainty. A common approach is leave-k-out (LkO) analysis, in which the model is fitted to datasets from which one or more observations have been removed (\cite{gelfand1992model}, \cite{peruggia1997variability}). If the exclusion of a particular observation leads to substantial changes in posterior estimates or credible intervals, that observation can be regarded as highly influential. Such sensitivity analyses help evaluate the extent to which inference depends on a small subset of the data. Classical bootstrap methods provide a Frequentist alternative for quantifying uncertainty and assessing the stability of statistical estimates (\cite{efron1994introduction}). Bootstrapping is a resampling-based algorithm that approximates the sampling distribution of an estimator, and is applicable even when the exact distribution of the estimator is analytically intractable. Unlike standard Frequentist approaches that rely on asymptotic assumptions, bootstrap methods can remain valid under relatively mild regularity conditions, provided that the empirical distribution reasonably approximates the true data-generating process. The resulting bootstrap distribution can be used to assess the bias and standard error of a point estimate, to construct confidence intervals, and to compute p-values for hypothesis testing. These two model evaluation approaches are  particularly appealing since applicable even when the model is misspecified.

However, these model evaluation procedures face difficulties when it is applied to Bayesian models that require MCMC methods to obtain posterior distributions. In such settings, the computational cost can be prohibitive since, for example, bootstrapping requires repeatedly resampling the observed data to construct multiple bootstrap datasets and refitting the same model to each dataset. In practice, this often involves 1,000 to 10,000 replications, leading to computational times that can range from several days to even weeks for complex hierarchical models (\cite{knape2018sensitivity}). Reducing the computational burden is desirable, particularly for Bayesian models in which posterior uncertainty estimates may be misleading. 

We focus on computationally efficient algorithms for obtaining model evaluation measures without requiring repeated refitting of the same Bayesian model. Recent developments in statistics have shown that several Frequentist properties of Bayesian estimators, including standard errors and confidence interval, can be approximated using influence functions derived from the posterior distribution. These influence functions can be expressed as posterior covariances involving the observation-level log-likelihood contributions (\cite{giordano2018covariances}; \cite{iba2025posterior}; \cite{ji2025valid}), making them straightforward to compute from posterior samples obtained in a single MCMC run. The underlying theory combines the infinitesimal jackknife (IJK) approximation (\cite{jaeckel1972infinitesimal}; \cite{giordano2019swiss}) with local case-sensitivity analysis (\cite{Gustafson}, \cite{perez2006mcmc}, \cite{millar2007assessment}, \cite {lesaffre2012bayesian}, \cite{giordano2018covariances}), establishing a connection between empirical influence functions and posterior sensitivity measures. While (\cite{Giordano2023ar}) focused primarily on the covariance formula, their framework naturally allows the construction of an approximate bootstrap algorithm; (\cite{iba2026w}) implemented such an algorithm and demonstrated its performance in simple models. An important feature of this framework is that it remains applicable even when the model is misspecified that it eliminates the need for repeated model refitting.

Despite these advantages, the methodology has seen limited adoption in applied research. One reason is that much of the existing literature emphasizes theoretical developments and mathematical properties, making the methodology less accessible to practitioners (e.g., \cite{Giordano2023ar}, \cite{iba2026w}). Second, its performance in more complex settings typical of ecological applications has been less investigated. Although a previous paper that proposed the approximated bootstrap has discussed its applicability to some hierarchical Bayesian models (\cite{iba2026w}), important classes of ecological models have not yet been examined, including discrete-outcome regression models (e.g. binary or count data) and hierarchical models with latent discrete parameters. These models are often essential for estimating site-occupancy indicators or latent population abundances. Finally, software implementations suitable for routine data analysis remain relatively scarce. 



In this paper, we examine the applicability of these methods to ecological data analysis. We first review the fundamental ideas underlying the framework and then introduce posterior-covariance sensitivity measures for identifying influential observations and approximating bootstrap distributions. Then, we conducted numerical experiments to evaluate their performance and demonstrate their practical utility through ecological case studies. The results of the numerical experiments show that the posterior-covariance approximations closely reproduce the results obtained by exact model refitting and provide better Frequentist coverage than Bayesian credible intervals, while requiring much less computation. 

To facilitate practical use, we also provide the \textsf{R} package \texttt{poscosea}, which implements sensitivity analysis and MCMC-based bootstrap approximations for a broad class of Bayesian models. Although the \texttt{StanSensitivity} package (\cite{StanSensitivity}) includes the
\texttt{ComputeIJCovariance} function for estimating Frequentist covariance, it does not provide a bootstrap approximation or the resulting interval estimation, and it is designed specifically for Stan. In contrast, \texttt{poscosea} provides leave-k-out sensitivity measures aimed at identifying influential observations, together with bootstrap-based uncertainty quantification. It is compatible with posterior samples obtained from any MCMC software, as long as posterior samples and observation-level log-likelihood values are available. 

The rest of the article is organized as follows. In the next section, we review the method with its basic rationale. We show that expanding the posterior expectation of a target quantity around the observation weight yields a covariance posterior formula for sensitivity measure. In section 3, we conduct numerical simulations to present how the method works for both sensitivity analysis and bootstrapping of Generalized Linear Mixed Model (GLMM). In section 4, we apply the method for N-mixture model to confirm the performance of the method in more complex Bayesian model with hierarchical structures. In section 5, we also test our methods to an empirical data originally analyzed by (\cite{kery2020applied}) and introduce the \textsf{R} package \texttt{poscosea} which implemented these methods in section 6.

\section{Methods}
\subsection{Common Settings and Notation}

In this paper, we work with a Bayesian approach using the posterior distribution:

\begin{equation}
\label{1}
p(\theta \mid X^n)
=
\frac{
\exp\!\left\{\sum_{i=1}^{n} s(X_i;\theta)\right\}p(\theta)
}{
\int
\exp\!\left\{\sum_{i=1}^{n} s(X_i;\theta')\right\}
p(\theta')\,d\theta'
},
\end{equation}

where \(p(\theta)\) is a prior density and
\(s(x;\theta)=\log p(x\mid\theta)\) is log-likelihood of the model. Our problem is to obtain Frequentist's properties of the posterior average

\[
E_{\mathrm{pos}}[A(\theta)]
=
\int A(\theta)\,p(\theta\mid X^n)\,d\theta
\]

of a given statistic \(A(\theta)\), where
\(X^n=(X_1,X_2,\ldots,X_n)\) is a sample from the true
distribution \(G\).
Here and hereafter, we denote a posterior average by
\(E_{\mathrm{pos}}\).
We also define \(\operatorname{Cov}_{\mathrm{pos}}\)
in a similar way.

For later convenience, we add another terms. First, a posterior distribution
\(p_w(\theta;X^n)\)
with Bayesian observation weight (BOW)
\((w_1,w_2,\ldots,w_n)\)
of the observations \(X^n=(X_1,X_2,\ldots,X_n)\)
is defined by

\begin{equation}
p_w(\theta;X^n)
=
\frac{
\exp\!\left\{
\sum_{i=1}^{n} w_i s(X_i;\theta)
\right\}
p(\theta)
}{
\int
\exp\!\left\{
\sum_{i=1}^{n} w_i s(X_i;\theta')
\right\}
p(\theta')\,d\theta'
}.
\end{equation}

The average of \(A(\theta)\) over the distribution
\(p_w(\theta;X^n)\) are expressed as

\begin{equation}
\label{E_posw}
E_{\mathrm{pos}}^{w}[A(\theta)]
=
\int A(\theta)\,p_w(\theta;X^n)\,d\theta .
\end{equation}

Here and hereafter \(w=1\) is used as an abbreviation of
\(w_j=1,\; j=1,2,\ldots,n\).

\subsection{Posterior covariance representation of sensitivity formula}
Based on these settings, we introduce a formula for BOW sensitivity proposed by (Giordano, 2020). See also (\cite{perez2006mcmc}; \cite{millar2007assessment}; Giordano et al., 2018; \cite{iba2025posterior}). 

Local case sensitivity of the posterior average $E_{pos}[A(\theta)]$ to the observation $x_i$ is defined as a derivative of $E_{pos}^w[A(\theta)]$ in (\ref{E_posw}) by the BOW $w_i$ evaluated at $w= 1$. The local case sensitivity formula expresses it by the posterior covariance between \(A\) and \(s(X_i;\theta)\) as

\begin{equation}
\label{deriv_w}
\frac{\partial}{\partial w_i}
E_{\mathrm{pos}}^{w}[A(\theta)]
=
\operatorname{Cov}_{\mathrm{pos}}
\!\left[
A,\,
s(X_i;\theta)
\right].
\end{equation}

The proof is straightforward, once an exchange of integration and derivative is allowed under appropriate regularity conditions.

\paragraph{Proof.}

By definition,
\begin{equation}
E_{\mathrm{pos}}^{w}[A(\theta)]
=
\frac{
\int
A(\theta)
\exp\!\left\{
\sum_{j=1}^{n}
w_j s(X_j;\theta)
\right\}
p(\theta)\,d\theta
}{
\int
\exp\!\left\{
\sum_{j=1}^{n}
w_j s(X_j;\theta)
\right\}
p(\theta)\,d\theta
}.
\end{equation}

Let

\[
L_w(\theta)
=
\exp\!\left\{
\sum_{j=1}^{n}
w_j s(X_j;\theta)
\right\}
p(\theta)
\]

and denote

\[
N(w)=\int A(\theta)L_w(\theta),d\theta,
\qquad
D(w)=\int L_w(\theta),d\theta.
\]

Then 
\[
E_{\mathrm{pos}}^{w}[A(\theta)]=N(w)/D(w)).
\]

Under regularity conditions that justify differentiation under the integral sign,

\begin{equation}
\frac{\partial N(w)}{\partial w_i} = \int A(\theta)s(X_i;\theta)L_w(\theta),d\theta,
\end{equation}

and

\begin{equation}
\frac{\partial D(w)}{\partial w_i}
=
\int s(X_i;\theta)L_w(\theta),d\theta.
\end{equation}

Thus,

\begin{align}
\frac{\partial}{\partial w_i}
E_{\mathrm{pos}}^{w}[A(\theta)]
&=
\frac{
E_{\mathrm{pos}}^{w}
[A,s(X_i;\theta)]
D(w)^2
-
E_{\mathrm{pos}}^{w}[A]
E_{\mathrm{pos}}^{w}[s(X_i;\theta)]
D(w)^2
}{
D(w)^2
}
\\[2ex]
&=
E_{\mathrm{pos}}^{w}
\!\left[A(\theta)s(X_i;\theta)\right]
-
E_{\mathrm{pos}}^{w}[A]
E_{\mathrm{pos}}^{w}[s(X_i;\theta)].
\end{align}

In practice, the posterior covariance in Eq.~(\ref{deriv_w}) can be readily estimated from posterior samples by computing the sample covariance between $A(\theta)$ and $s(X_i;\theta)$.

\subsection{Leave-k-out Influence Measures}
Utilizing this result, we show that the change in the posterior mean estimate $E_{\mathrm{pos}}[A(\theta)]$ when one or more observations are removed from the dataset can be approximated in a linear form. Suppose the posterior draws were generated by MCMC or anything approximating methods for the exact posterior distribution $p(\theta|X^n)$ by which posterior mean $E_{\mathrm{pos}}[A(\theta)]$ is calculated. Then, assuming $E_{\mathrm{pos}}^{w}[A(\theta)]$ is a smooth function of the BOW ($w=(w_1,\ldots,w_n)$), a first-order Taylor expansion around ($w=1$) yields

\begin{align}
E_{\mathrm{pos}}^{w}[A(\theta)]
&\approx
E_{\mathrm{pos}}[A(\theta)]
+
\sum_{i=1}^{n}
(w_i-1)
\left.
\frac{\partial}{\partial w_i}
E_{\mathrm{pos}}^{w}[A(\theta)]
\right|_{w=1} \\
&=
E_{\mathrm{pos}}[A(\theta)]
+
\sum_{i=1}^{n}
(w_i-1)
\operatorname{Cov}_{\mathrm{pos}}
\!\left[
A,
s(X_i;\theta)
\right]
\label{eq:loo_approx}
\end{align}

Removing an observation from the dataset corresponds to setting its BOW to zero. More generally, deleting a subset of observations can be represented by replacing the corresponding elements of $w$ with zero while keeping the remaining BOW equal to one. Since $E_{\mathrm{pos}}[A(\theta)]$ can be obtained from the posterior draws from the original data, we can assess which observations have the largest influence on the posterior estimate and quantify the magnitude of their impact without re-fitting the model, by repeatedly evaluating the above approximation under different choices of $w$.

\subsection{Bootstrap Approximation}
The posterior covariance representation above  can be further extended to emulate bootstrap resampling without additional MCMC runs (e.g. \cite{iba2026w}). While (\cite{Giordano2023ar}) and (\cite{luo2026robust}) discussed closed-form expressions for Frequentist covariance, the bootstrap-approximation approach adopted here has the practical advantage of being applicable to a wide range of settings without requiring problem-specific analytical derivations. As removing the $i-$th observation corresponds to setting its BOW to ($w_i=0$), a bootstrap replicate in which the same observation is resampled $R_i$ times corresponds to assigning the BOW $w_i=R_i$. Therefore, a bootstrap sample can be represented by a vector of BOW, allowing the effect of resampling to be approximated through a Taylor expansion of the posterior mean with respect to $w$ (see also \cite{efron1994introduction}; \cite{giordano2019swiss}).

Suppose that bootstrap resampled data are generated $B$ times. For each replicate ($b=1,\ldots,B$), draw multinomial counts $R_i^{(b)}; i=1,\ldots,n$, according to

\begin{equation}
\label{weight}
(R_1^{(b)},R_2^{(b)},\ldots,R_n^{(b)})
\sim
\mathrm{Multinomial}
\left(
\frac1n,\frac1n,\ldots,\frac1n
\right),
\end{equation}

which characterizes how many times $i-$th observation is resampled from the original data in $b-$th bootstrapped data. Utilizing Eq.(\ref{deriv_w}) to represent the derivatives, we obtain the approximated posterior mean of $A(\theta)$ under a $b-$th bootstrapped data $E_{\mathrm{pos}}[A(\theta)^{(b)}]$ as

\begin{equation}
\label{Boot_A}
E_{\mathrm{pos}}[A(\theta)^{(b)}]
=
E_{\mathrm{pos}}[A]
+
\sum_{i=1}^{n}
(R_i^{(b)}-1)
\operatorname{Cov}_{\mathrm{pos}}
\left[
A(\theta),
s(X_i;\theta)
\right].
\end{equation}

Repeating this process $B$ times results in a collection of approximated posterior mean values, which correspond to the bootstrap distribution of the posterior mean estimates. The approximated method requires only the generation of multinomial counts $(R_1^{(b)},\ldots,R_n^{b})$ to approximate bootstrap resampling; no additional MCMC runs are needed. In contrast to (\cite{lee2017frequentist}), which approximated the bootstrap using importance sampling, the present method tends to be faster when the number of bootstrap samples is large and to be less sensitive to the particular posterior-sample realization (\cite{iba2026w}). Consequently, it can substantially reduce the computational burden associated with conventional bootstrap procedures. 

\section{A Toy Example: Generalized Linear Mixed Model under a Misspecification}
We conducted numerical simulations to evaluate the performance of the method. For each simulated dataset, count data $y_i$, $i=1,2,...n$ were generated according to the following over-dispersed logistic regression model with the number of trials $N$:

\begin{align}
\label{DGP1}
y_i &\sim \mathrm{Binomial}(N,p_i), \\
\label{DGP2}
    p_i &\sim \mathrm{beta}(\kappa \mu_i,(1-\mu_i)\kappa), \\
\label{DGP3}
\operatorname{logit}(\mu_i) &= x_i^T\beta + z_i^T b, \\
\label{DGP4}
b &\sim \mathcal{N}(0,I).
\end{align}

where $x_i$ and $z_i$ denote the $i-$th rows of the fixed- and random-effects design matrices X and Z, respectively. The fixed-effects design matrix consisted of an intercept and a covariate independently generated from a uniform distribution on [-1,1]. The random-effects design matrix Z encoded site-specific intercepts. The parameter $\kappa$ controls the degree of over-dispersion in the Beta distribution, with smaller values corresponding to greater variability in the probability. The true parameter values were fixed at $N=20, \beta=(0.25,\ 0.5)^\top$, and $\kappa=5$.

We applied the following misspecified model that dismisses the over-dispersion term with priors:
\begin{align}
y_i
&\sim
\mathrm{Binomial}(N, p_i), \\
\operatorname{logit}(p_i)
&=
x_i^\top\beta + z_i^\top b, \\
\beta
&\sim
\mathcal N(0,10^6I), \\
b
&\sim
\mathcal N(0, I).
\end{align}

We implemented Polya–Gamma Gibbs sampler for drawing samples from the posterior distribution. Simulation experiments were repeated for each sample size ($n \in {50, 100,200}$), where the number of groups for random-effect was $5$. All analyses were conducted in \textsf{R} (version 4.3).

\subsection{Sensitivity Analysis}
\begin{figure}[tbh]
\centering
\includegraphics[width=15cm]{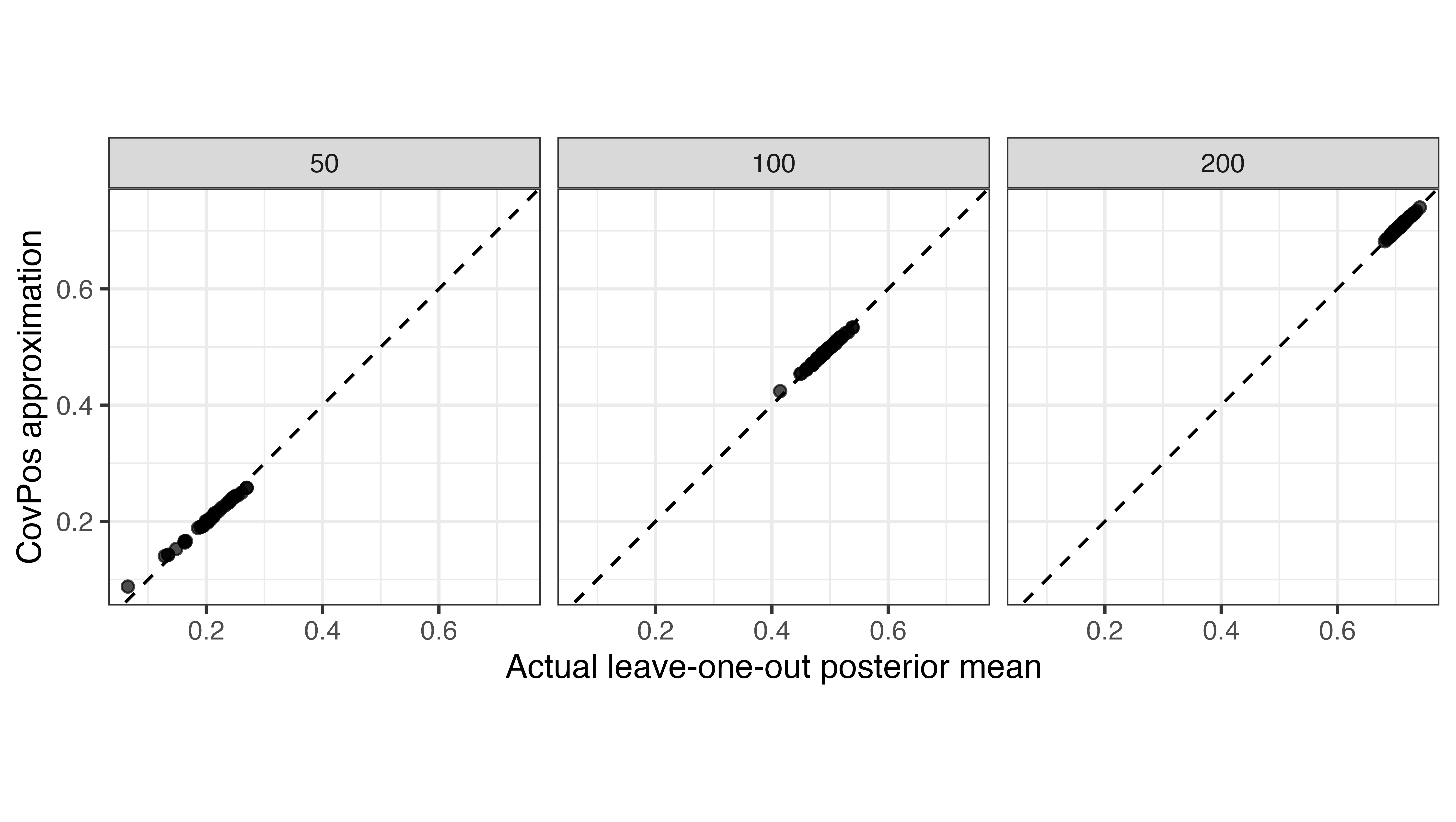}
\caption{
The scatter plot for comparing the actual posterior mean estimates obtained via the re-fitting of the same model and those of the approximated estimates by the posterior-covariance method ($n \in \{50, 100,200\}$).}
\label{fig:1}
\end{figure}
We compared two approaches for quantifying the influence of individual observations: (1) the actual refitting of the same model without a specific observation and (2) approximated posterior change obtained by the posterior-covariance measure. For each simulation replicate, data were generated according to (\ref{DGP1} to \ref{DGP4}), and 10,000 posterior samples were drawn, with the first 2,000 samples discarded as burn-in. Using the remaining samples, we calculated the posterior mean of $\beta$ and the log-likelihood of each observation. These quantities were then used to compute the posterior-covariance measure for every observation ($i=1,\ldots,n$). To obtain a benchmark measure of influence, we performed an additional leave-one-out analysis. For each observation $i$, we refitted the same model after removing that observation from the dataset and recorded the resulting change in the posterior mean of $\beta$. This procedure was repeated for all observations in the dataset. 

Fig. (\ref{fig:1}) illustrates a representative realization of the result of one trial. Comparing the actual change in the posterior mean obtained from leave-one-out refitting with the corresponding posterior-covariance approximated measure for each observation ($i=1,\ldots,n$), a strong correlation is confirmed for all three cases. 

\begin{table}
\caption{
Mean Spearman rank correlations (standard errors in parentheses) over 100 simulation replicates between the actual leave-one-out change in the posterior mean and the posterior-covariance measure. For each simulation replicate, the correlation was computed across all observations in the dataset, comparing the observed leave-one-out changes with the corresponding influence scores ($n=50$, $100$, and $200$). Larger correlations of the posterior-covariance method indicate better agreement with the true influence of observations on posterior inference and therefore greater accuracy in identifying influential observations.
}
\label{tab1}
\centering
\begin{tabular}[t]{ccc}
\toprule
Sample Size  & Spearman rank correlation\\
\midrule
50 & 0.978 (0.002)\\
100 & 0.982 (0.001)\\
200 & 0.983 (0.001)\\
\bottomrule
\end{tabular}
\end{table}

Table \ref{tab1} summarizes the results across the 100 simulation replicates. For each replicate, we calculated the Spearman rank correlation across the $n$ observations between the actual leave-one-out change in the posterior mean and each influence measure. The posterior-covariance measure exhibited correlations with the true changes in the posterior mean. These results suggest that the posterior-covariance method provides a useful approximation for identifying observations that strongly influence parameter estimates and for quantifying the magnitude of the resulting change in posterior inference.

\subsection{Bootsrtap}
We also tested the performance of the methods for approximating the bootstrapping without re-fitting to the resampled data. As was tested in Sensitivity Analysis, data were generated according to (\ref{DGP1} to \ref{DGP4}) and obtained posterior mean of $\beta$ with log-likelihood. Then, we applied the algorithm with $B=5000$ bootsrtap replications. This process was repeated 1,000 times, and we evaluated the Frequentist standard error and the coverage rate (i.e., the proportion of times the 95 percent interval for a parameter contained the true value).

Table \ref{tab:sim_glmm} summarizes the results of the simulation study. First, we found that the Bayesian posterior distribution underestimated the uncertainty of the estimated parameter, when viewed from the Frequentist account. The posterior standard deviation was consistently smaller than the empirical standard deviation of the posterior mean estimates across the simulated datasets. Consequently, the Frequentist coverage of the Bayesian $95\%$ credible interval fell below the nominal level, indicating that the posterior credible interval may provide an optimistic assessment of uncertainty when used as an alternative to Frequentist confidence interval. In contrast, the bootstrap approximation substantially improved the uncertainty assessment, yielding standard errors and confidence intervals that were much closer to their empirical Frequentist ones.

\begin{table}[t]
\centering
\caption{Comparison of Bayesian posterior uncertainty and bootstrap approximation. For each sample size, uncertainty of the point estimates (left) and the coverage performance of interval estimates (right) are presented. The results indicate that the posterior quantification of uncertainty is too optimistic from Frequentist account, while the approximated bootstrapping offers a more reliable ones.}
\label{tab:sim_glmm}
\begin{tabular}{lccc|cc}
\toprule
&
\multicolumn{3}{c|}{Standard error}
&
\multicolumn{2}{c}{95\% coverage}
\\
\cmidrule(lr){2-4}
\cmidrule(l){5-6}
Sample Size
&
Empirical&
Mean of Posterior SD
&
Bootstrap
&
Posterior
&
Bootstrap
\\
\midrule
50  &
0.285 &
0.133 &
0.252 &
0.646 &
0.913 \\

100 &
0.186 &
0.091 &
0.178 &
0.663 &
0.933 \\

200 &
0.129 &
0.063 &
0.126 &
0.680 &
0.946 \\
\bottomrule
\end{tabular}
\end{table}

\section{Application to the N-mixture Model}
To demonstrate the applicability of the method to more complex Bayesian hierarchical models, we applied it to an N-mixture model. Because the N-mixture model involves latent abundance variables and nonlinear hierarchical components, this application can be a realistic test case for evaluating the practical usefulness of the method in ecology beyond simple generalized linear mixed models.

For the $j$-th survey ($j=1,2,...J$) of $i$-th population ($i=1,2,...I$), let $N_i$ denote the latent abundance and $y_{ij}$ the observed count. A typical N-mixture model is specified as

\begin{align}
y_{ij}
    &\sim \mathrm{Binomial}(N_i, p_{ij}), \\
N_i
    &\sim \mathrm{Poisson}(\lambda_i), \\
\log (\lambda_i) &= \alpha_{\lambda} + x_i^\top \beta_{\lambda}, \\
\operatorname{logit}(p_{ij})
    &= \alpha_{p} + z_{ij}^\top \beta_{p}.
\end{align}

Here, $x_i$ is a vector of site-level covariates affecting abundance, and $\alpha_{\lambda}, \beta_{\lambda}$, is the corresponding regression coefficients. The latent abundance $N_i$ is assumed to follow a Poisson distribution with mean $\lambda_i$. The detection probability $p_{ij}$ represents the probability of detecting an individual during survey $j$ at site $i$, where $z_{ij}$ denotes a vector of observation-level covariates and $\alpha_p, \beta_p$ is the associated coefficients. The model is often fitted using Bayesian inference, and posterior samples are obtained by MCMC. 

We evaluated the posterior-covariance methods for estimating the abundance parameters $\alpha_{\lambda}$ and $\beta_{\lambda}$, as well as the total population size of the surveyed area, $A(\theta)=\sum_{i=1}^I N_i$, under three different model misspecification scenarios. (a) Gaussian overdispersion, $r_i \sim \mathcal{N}(0,\sigma^2)$ wtih $\sigma=0.5$, was added to the linear predictor of $\lambda_i$; (b) the latent abundance $N_i$ was generated from a negative binomial distribution instead of a Poisson distribution as $V[N_i]=\lambda_i+ \phi\lambda_i^2$, with $\phi=0.1$ while preserving the same expected abundance, $E[N_i]=\lambda_i$; and (c) the detection probability was generated from a beta distribution, $p_{it}\sim \mathrm{Beta}(\kappa\mu_{it},(1-\mu_{it})\kappa)$, with $\kappa=10$, rather than being fixed conditional on the covariates. The covariates $x_i$ and $z_{ij}$ were independently generated from the standard normal distribution,$ \mathcal{N}(0,1)$. The true parameter values were fixed at $\alpha_{\lambda}=2.0, \beta_{\lambda}=-2.0, \alpha_p=-0.2$,  $\beta_p=0.5$.

In contrast, inference was performed using the misspecified latent Poisson model that ignores the over-dispersion component with weakly informative priors assigned to all regression coefficients:

\begin{align}
\alpha_{\lambda}
&\sim
\mathcal N(0,10^3), \\
\beta_{\lambda}
&\sim
\mathcal N(0,10^3), \\
\alpha_{p}
&\sim
\mathcal N(0,10^3), \\
\beta_{p}
&\sim
\mathcal N(0,10^3).
\end{align}

We considered three sample sizes, $I \in {50, 100,150}$ with $J=3$. We first obtained 10000 posterior draws by MCMC, where the first 5000 samples were discarded as burn-in. Then, we calculated its convergence measure \(\hat{R}\) to ensure the reliability of the posterior inference. If $ \hat{R} > 1.01$, we continued to drawing more posterior samples until  $\hat{R} < 1.01$. The posterior-covariance measure were then computed for each observation.

In Bayesian hierarchical models with latent variables, the choice of log-likelihood used for posterior sensitivity analysis plays an important role. In general, the appropriate likelihood depends on the inferential target and the observational unit of interest (see also \cite{millar2018conditional,gaya2024comparison}). For inference on the abundance model parameters $\alpha_{\lambda}$ and $\beta_{\lambda}$, we used the joint log-likelihood, treating the latent abundances $N_i$ as model parameters. This choice is required because these parameters describe the conditional abundance process given the latent states, and the posterior sensitivity should therefore be evaluated with respect to the complete hierarchical model. In contrast, for inference on the total abundance $A(\theta)=\sum_{i=1}^n N_i$, we used the marginal log-likelihood, obtained by integrating out the latent abundance variables $N_i$. Since the inferential target is the population abundance itself rather than the latent states at individual sites, the leave-one-out perturbation should reflect the predictive contribution of each observed site after accounting for uncertainty in $N_i$. For each site, the marginal likelihood was computed by numerically summing over feasible values of $N_i$. Preliminary investigations confirmed that the probability mass beyond the specified truncation point $\text{max}(y_{i1}, ...y_{iJ_i})+100$ was negligible for simulated datasets considered.

The performance of approximated leave-k-estimate was evaluated by comparing them with the actual change in the posterior mean obtained by re-fitting the model after removing each $k=1$ observation. Since the model has a hierarchical structure of data, we removed whole data of $i=1,\ldots,I$ population while keeping the data of $j=1,\ldots,J$.

For the approximated bootstrap with the replications $B=5000$, we compared the empirical standard deviation of the posterior mean across simulated datasets, the average posterior standard deviation, the average bootstrap-estimated standard deviation, and the empirical coverage probabilities of the corresponding $95\%$ interval estimates. The simulation experiment was repeated 100 times for each sample size and for each misspecified scenarios. All analyses were conducted in \textsf{R} (version 4.3) using the packages \texttt{rjags} for sampling from the posterior distribution.

\begin{figure}[p]
\centering

\subcaptionbox{Log-normal scenario}{
\includegraphics[width=0.74\linewidth]{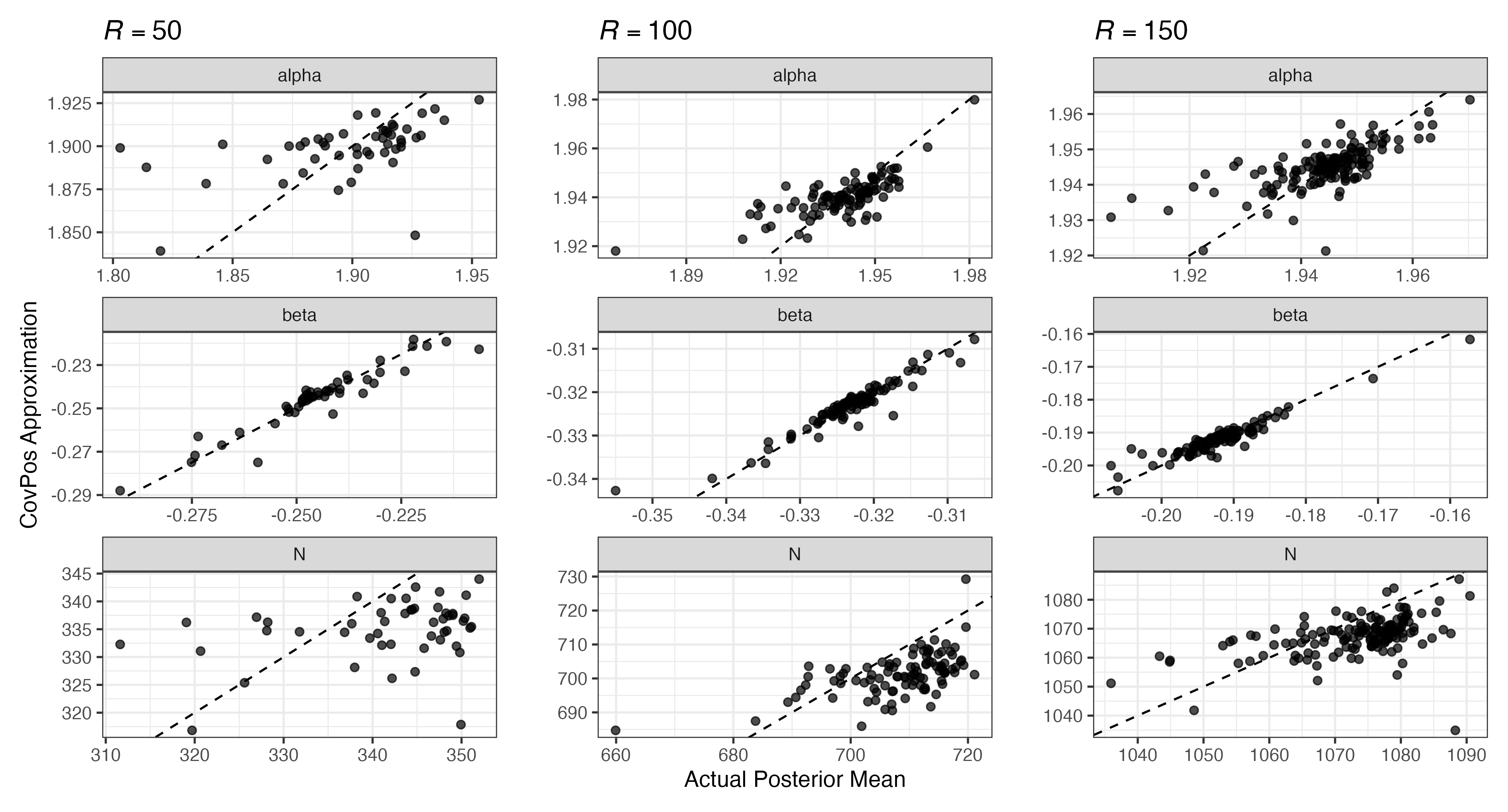}
}

\subcaptionbox{Negative binomial scenario}{
\includegraphics[width=0.74\linewidth]{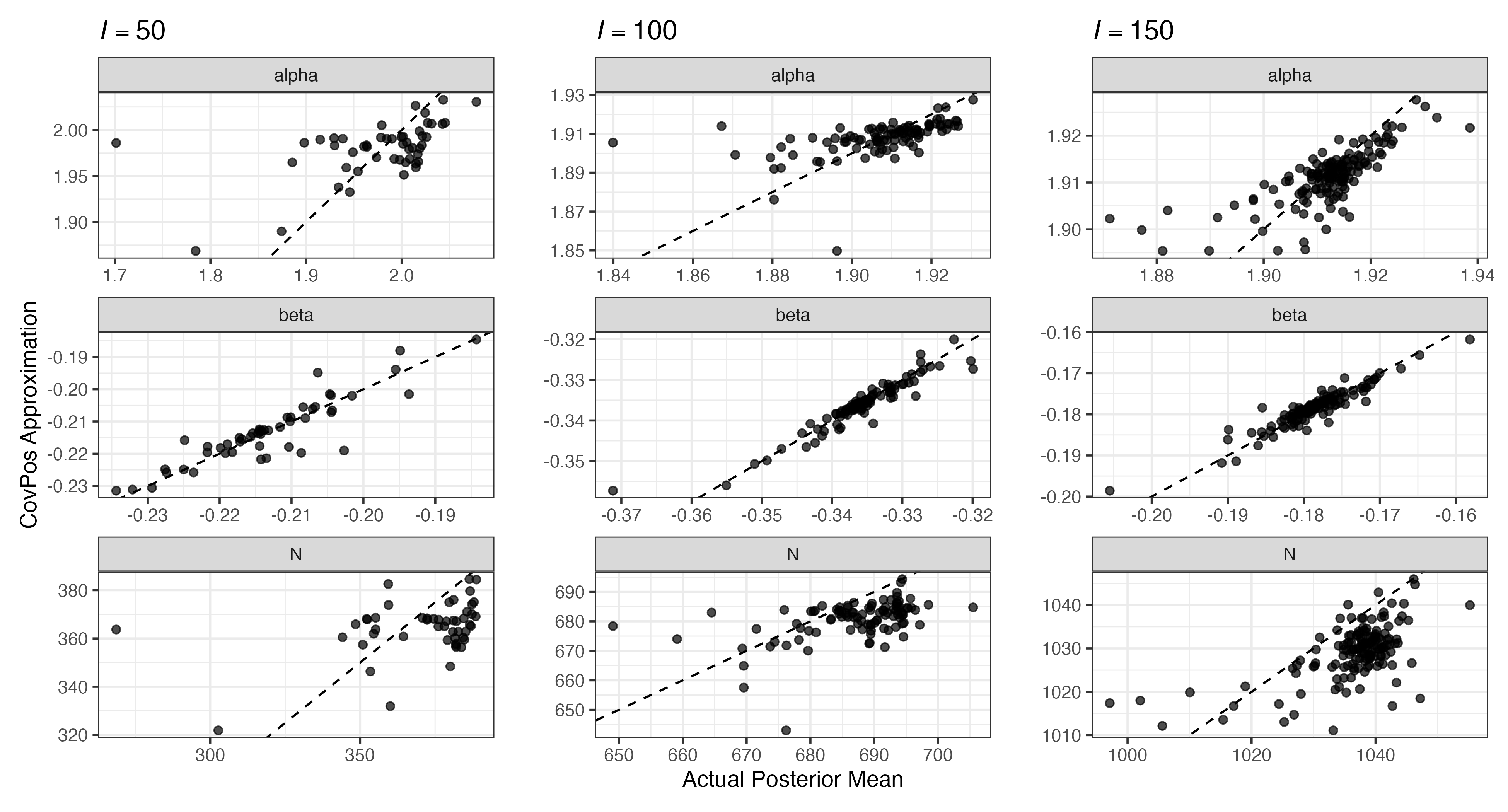}
}

\subcaptionbox{Beta-binomial scenario}{
\includegraphics[width=0.74\linewidth]{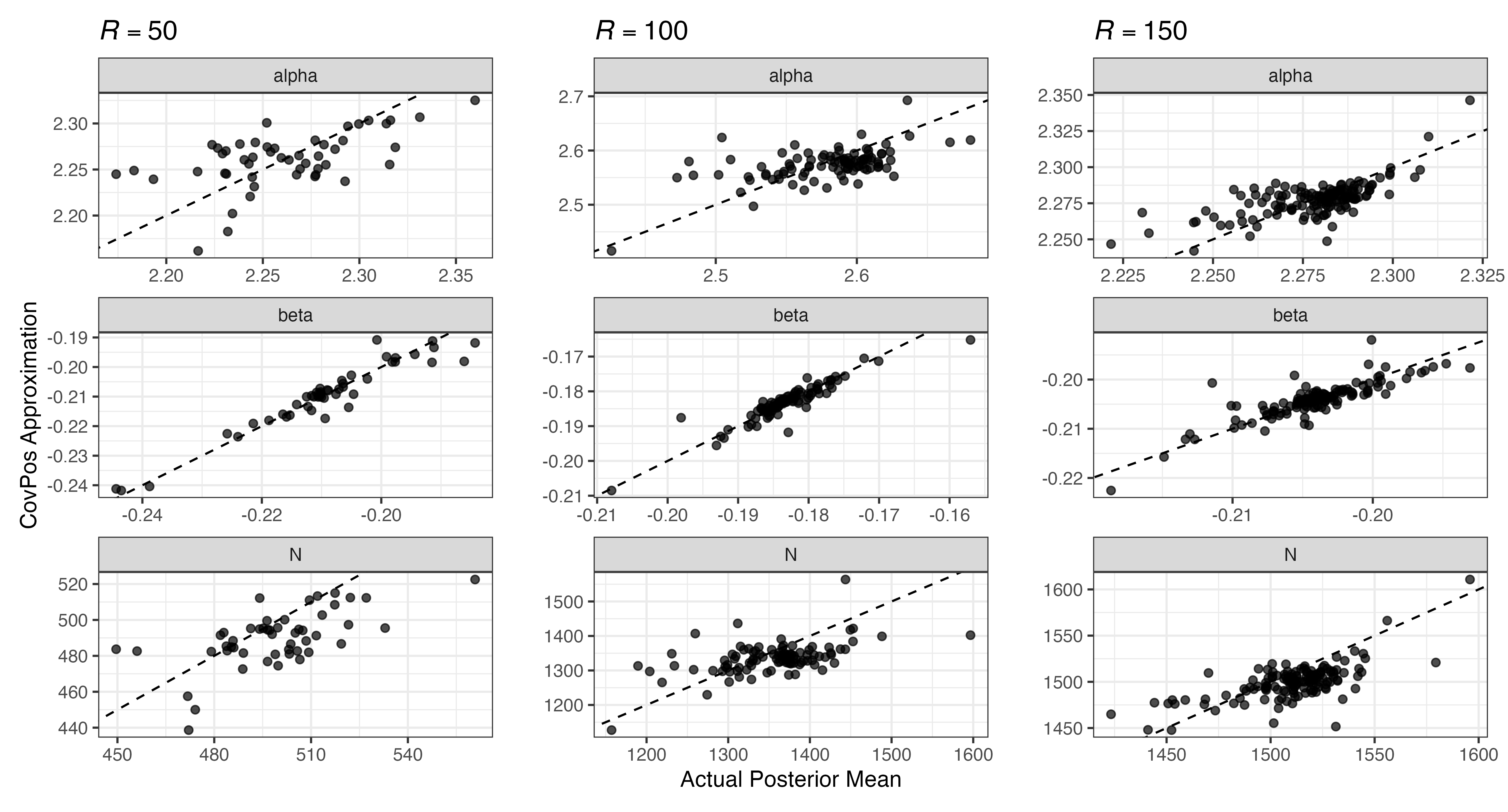}
}

\caption{
The scatter plots comparing the actual posterior mean estimates obtained by re-fitting the model with the approximated estimates from the posterior-covariance method ($I\in\{50,100,150\}$). Panels (a)--(c) correspond to the log-normal, negative-binomial, and beta-binomial misspecification scenarios, respectively.
}
\label{fig:2}
\end{figure}

Figure \ref{fig:2} compares the leave-one-out posterior mean obtained by repeatedly refitting the model with that approximated by the posterior-covariance method. Consistent with the results for the GLMM, the approximated posterior mean showed linear agreement with the leave-one-out estimates, indicating that it captures the influence of individual observations without requiring additional MCMC runs. Although the agreement is weaker than in the GLMM case, particularly for $\alpha_{\lambda}$ and $\sum_i^I N_i$, the greater variability observed in the N-mixture results may partly reflect Monte Carlo error in the MCMC estimates. The N-mixture model requires a Metropolis-based sampling scheme, whereas the GLMM considered above can be efficiently sampled using a Gibbs sampler. Consequently, the posterior quantities obtained from the N-mixture model may be subject to greater Monte Carlo error. Increasing the number of MCMC iterations could therefore improve the agreement between the exact refitting and the approximation, although the contribution of MCMC error to the observed variability remains to be investigated.

\begin{table}[!h]
\centering
\caption{Comparison of Bayesian posterior uncertainty and bootstrap approximation for the N-mixture model under the three different misspecified scenarios. For each sample size, uncertainty of the point estimates (left) and the coverage performance of interval estimates (right) are presented. The results indicate that the posterior quantification of uncertainty is too optimistic from Frequentist account, while the approximated bootstrapping offers a more reliable ones.}
\label{tab:sim_nmix}
\begin{subtable}{\linewidth}
\centering
\caption{log-Normal Misspecification}
\begin{tabular}{llccccc}
\toprule
\multicolumn{2}{c}{ } & \multicolumn{3}{c}{Standard error} & \multicolumn{2}{c}{95\% coverage} \\
\cmidrule(l{3pt}r{3pt}){3-5} \cmidrule(l{3pt}r{3pt}){6-7}
Sample Size & Parameter & Empirical & Mean of Posterior SD  & Bootstrap & Posterior & Bootstrap\\
\midrule
 & $\alpha_{\lambda}$ & 0.10 & 0.10 & 0.16 & 0.90 & 0.98\\
\cmidrule{2-7}
 & $\beta_{\lambda}$ & 0.09 & 0.06 & 0.10 & 0.80 & 0.95\\
\cmidrule{2-7}
\multirow{-3}{*}{\raggedright 50} & $\sum N$ & 46.28 & 45.01 & 69.05 & 0.90 & 0.94\\
\cmidrule{1-7}
 & $\alpha_{\lambda}$ & 0.08 & 0.07 & 0.12 & 0.86 & 0.97\\
\cmidrule{2-7}
 & $\beta_{\lambda}$ & 0.06 & 0.04 & 0.07 & 0.81 & 0.97\\
\cmidrule{2-7}
\multirow{-3}{*}{\raggedright 100} & $\sum N$ & 67.64 & 58.53 & 91.39 & 0.85 & 0.91\\
\cmidrule{1-7}
 & $\alpha_{\lambda}$ & 0.07 & 0.05 & 0.09 & 0.79 & 0.96\\
\cmidrule{2-7}
 & $\beta_{\lambda}$ & 0.05 & 0.03 & 0.06 & 0.84 & 1.00\\
\cmidrule{2-7}
\multirow{-3}{*}{\raggedright 150} & $\sum N$ & 81.73 & 68.45 & 105.57 & 0.76 & 0.92\\
\bottomrule
\end{tabular}
\end{subtable}

\begin{subtable}{\linewidth}
\centering
\caption{Negative Binomial Misspecification}
\begin{tabular}{llccccc}
\toprule
\multicolumn{2}{c}{ } & \multicolumn{3}{c}{Standard error} & \multicolumn{2}{c}{95\% coverage} \\
\cmidrule(l{3pt}r{3pt}){3-5} \cmidrule(l{3pt}r{3pt}){6-7}
Sample Size & Parameter & Empirical & Mean of Posterior SD & Bootstrap & Posterior & Bootstrap\\
\midrule
 & $\alpha_\lambda$ & 0.15 & 0.13 & 0.21 & 0.84 & 0.95\\
\cmidrule{2-7}
 & $\beta_{\lambda}$ & 0.07 & 0.06 & 0.08 & 0.90 & 0.98\\
\cmidrule{2-7}
\multirow{-3}{*}{\raggedright 50} & $\sum N$ & 61.31 & 47.20 & 71.37 & 0.79 & 0.86\\
\cmidrule{1-7}
 & $\alpha_\lambda$ & 0.09 & 0.09 & 0.14 & 0.83 & 0.97\\
\cmidrule{2-7}
 & $\beta_\lambda$ & 0.05 & 0.04 & 0.06 & 0.89 & 0.96\\
\cmidrule{2-7}
\multirow{-3}{*}{\raggedright 100} & $\sum N$ & 68.03 & 56.88 & 86.99 & 0.79 & 0.93\\
\cmidrule{1-7}
 & $\alpha_\lambda$ & 0.08 & 0.07 & 0.11 & 0.75 & 0.96\\
\cmidrule{2-7}
 & $\beta_\lambda$ & 0.04 & 0.04 & 0.05 & 0.88 & 0.95\\
\cmidrule{2-7}
\multirow{-3}{*}{\raggedright 150} & $\sum N$ & 83.43 & 65.19 & 99.80 & 0.69 & 0.86\\
\bottomrule
\end{tabular}
\end{subtable}

\begin{subtable}{\linewidth}
\centering
\caption{Beta-Binomial Misspecification}
\begin{tabular}{llccccc}
\toprule
\multicolumn{2}{c}{ } & \multicolumn{3}{c}{Standard error} & \multicolumn{2}{c}{95\% coverage} \\
\cmidrule(l{3pt}r{3pt}){3-5} \cmidrule(l{3pt}r{3pt}){6-7}
Sample Size & Parameter & Empirical & Mean of Posterior SD & Bootstrap & Posterior & Bootstrap\\
\midrule
 & $\alpha_\lambda$ & 0.18 & 0.17 & 0.27 & 0.21 & 0.88\\
\cmidrule{2-7}
 & $\beta_\lambda$ & 0.04 & 0.04 & 0.05 & 0.95 & 0.96\\
\cmidrule{2-7}
\multirow{-3}{*}{\raggedright 100} & $\sum N$ & 217.10 & 237.68 & 569.37 & 0.20 & 1.00\\
\cmidrule{1-7}
 & $\alpha_\lambda$ & 0.15 & 0.14 & 0.22 & 0.12 & 0.73\\
\cmidrule{2-7}
 & $\beta_\lambda$ & 0.04 & 0.03 & 0.04 & 0.90 & 0.94\\
\cmidrule{2-7}
\multirow{-3}{*}{\raggedright 150} & $\sum N$ & 251.22 & 234.25 & 379.85 & 0.07 & 0.96\\
\bottomrule
\end{tabular}
\end{subtable}
\end{table}

Table \ref{tab:sim_nmix}. summarizes the results of the bootstrap simulation for the N-mixture model. For the beta binomial scenario, we report only the results of $I \in 100, 150$ because MCMC did not converge for some simulated data set of $I \in 50$. Consistent with the GLMM results, the posterior sd tends to be smaller than empirical one, underestimating the uncertainty of the estimated parameters. Consequently, the $95\%$ credible intervals failed to achieve the intended coverage probability. In contrast, the bootstrap approximation provided substantially more reliable uncertainty estimates, yielding confidence intervals with improved Frequentist coverage.

\section{Real Data Analysis}
To illustrate its practical utility, we reanalyzed an empirical dataset studied by (\cite{kery2020applied}). 
They considered the abundance of Swiss tit species using data of field surveys in Switzerland between 2004 and 2013. Then, they fitted several N-mixture models and estimated the abundances of them as quadratic functions of elevation. The resulting abundance estimates were subsequently used to investigate ecological patterns along elevational gradients and the national wide total abundance.

Using the same data, we fitted the following abundance model with three site-level covariates, elev, forest, and iRoute, for sites $i=1,\ldots,263$:

\begin{align}
\log(\lambda_i)
&=
\beta_0
+ elev_i \beta_1
+ elev_i^2 \beta_2
+ forest_i \beta_3
+ forest_i^2 \beta_4 \\
&\quad
+ elev_i\,forest_i^2 \beta_6
+ iRoute_i \beta_7, \notag\\
N_i
&\sim \mathrm{Poisson}(\lambda_i).
\end{align}

The detection model included 13 covariates. Although the original study employed a zero-inflated Poisson model for abundance, we intentionally used a standard Poisson model because their results exhibited some degree of over-dispersion, providing a test case for comparing the performance of the methods under model misspecification. The MCMC settings were the same as those used in the previous study, except that the model was implemented in \texttt{JAGS} rather than \texttt{WinBUGS}.

We obtained the marginal posterior regression curve of the expected abundance conditional on elevation as

\[
A(\theta; elev) =\exp (\beta_0
+ elev \beta_1
+ elev^2 \beta_2).
\]

\begin{figure}[tbh]
\centering
\includegraphics[width=15cm]{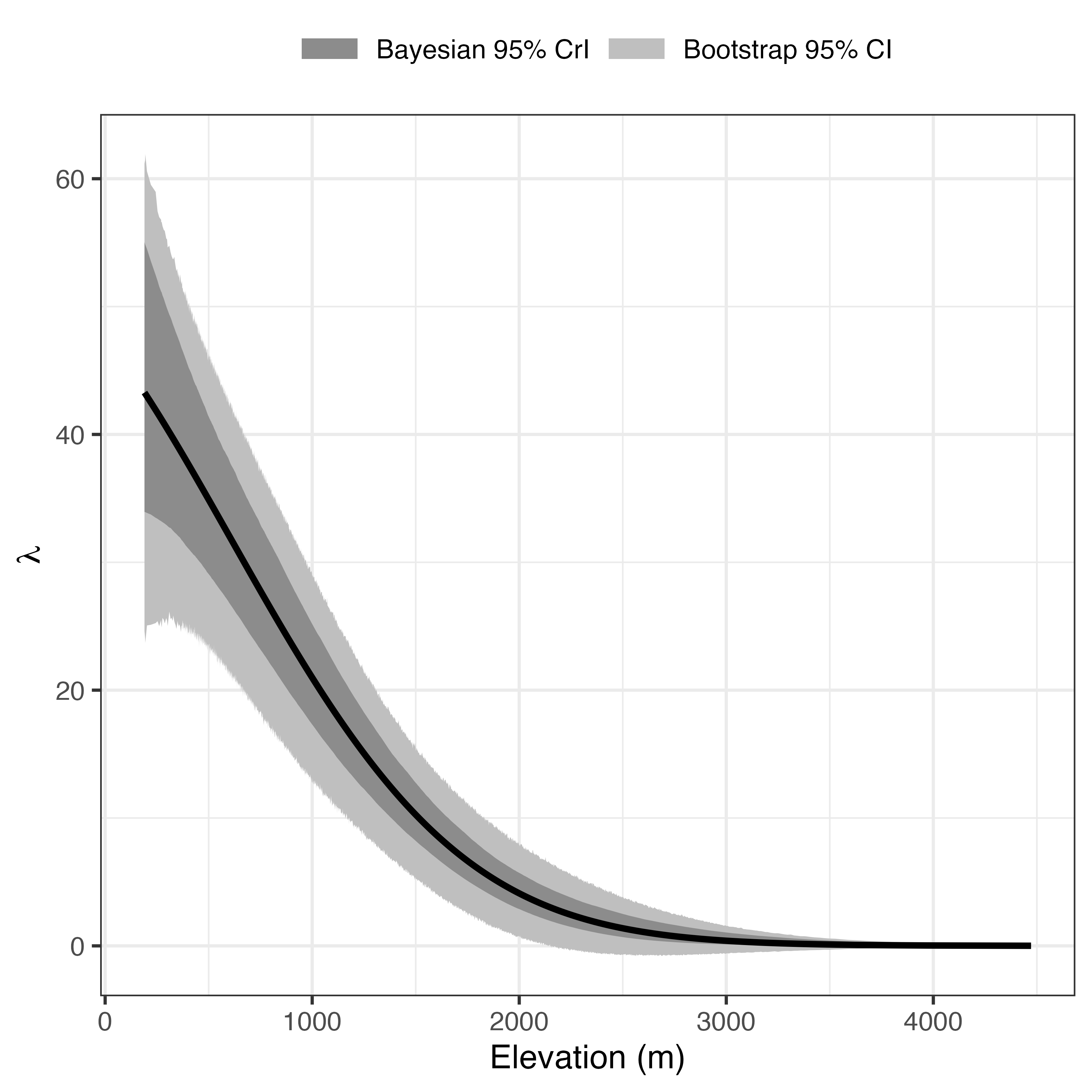}
\caption{
Estimated abundance–elevation relationship based on the posterior mean of $\lambda_{\mathrm{pred}}$. Dark and light gray bands denote the Bayesian $95\%$ credible interval and the $95\%$ confidence interval obtained from the bootstrap approximation, respectively. The posterior-covariance method yields wider uncertainty intervals, particularly at lower elevations where uncertainty in abundance estimates is greatest.}
\label{fig:apli}
\end{figure}

\begin{figure}[tbh]
\centering
\includegraphics[width=15cm]{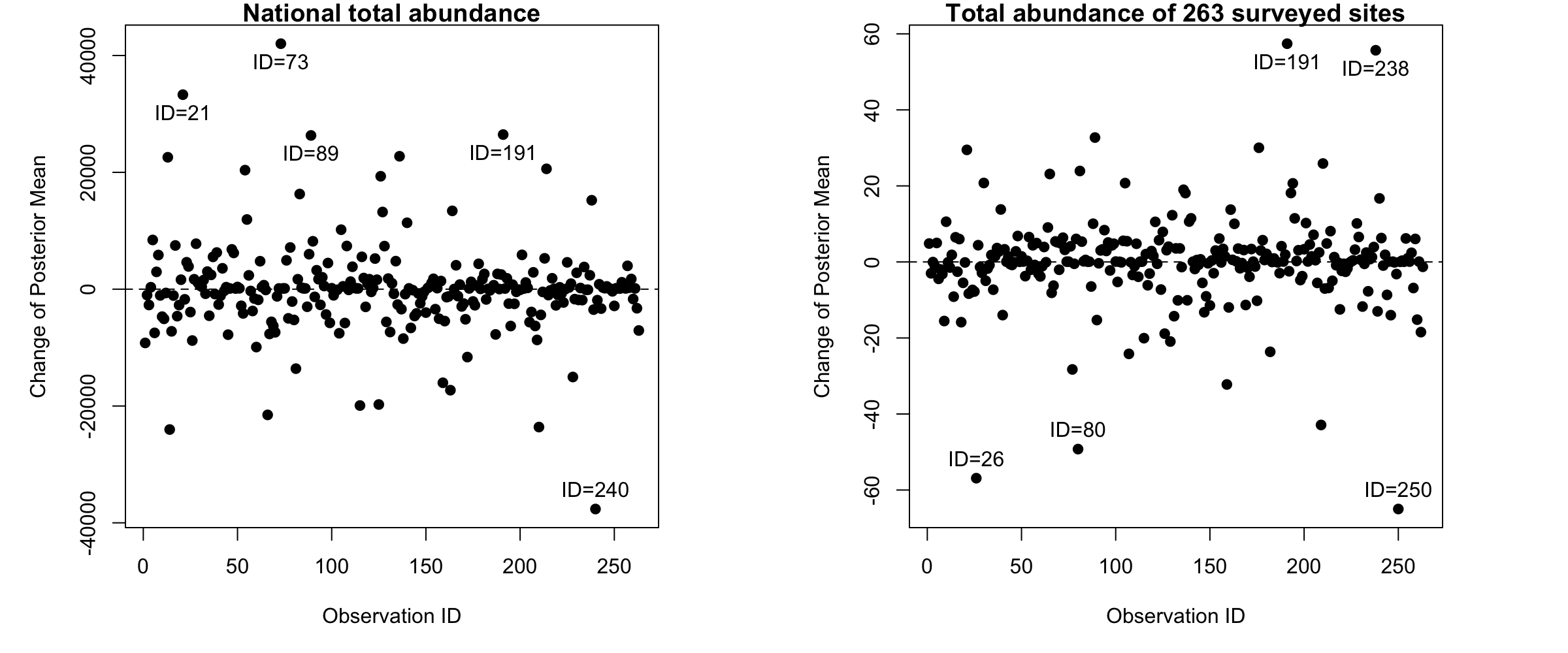}
\caption{
Leave-one-out diagnostic for two posterior quantities. The left panel shows the approximated change in the posterior mean of the estimated total national abundance, whereas the right panel shows the corresponding change for the total abundance across the 263 surveyed sites, after removing each sampling site from the analysis. The IDs of the five most influential sampling sites (largest absolute percentage changes) are labeled in each panel. The influential sites differ between the two target quantities, illustrating that observation-level influence depends on the posterior quantity of interest.}
\label{fig:LOO}
\end{figure}

Figure \ref{fig:apli} shows the posterior mean, $E_{\mathrm{pos}}[\lambda_{\mathrm{pred}}]=E_{\mathrm{pos}}[A(\theta; elev)] $, together with uncertainty bands based on the Bayesian $95\%$ credible interval and the $95\%$ confidence interval obtained using the bootstrap approximation. The confidence intervals produced by the posterior-covariance method were consistently wider than the corresponding Bayesian credible intervals. The discrepancy was particularly clear at lower elevations, where the abundance estimates exhibited greater uncertainty. These results suggest that the posterior-covariance method captures an additional source of variability that is not reflected in the posterior credible intervals, presumably owing to overdispersion in the Poisson abundance or the heterogeneity in the logistic detection probabilities.

The estimated national population size, $A(\theta)=\sum \lambda_{\mathrm{pred}}$, was \(7.38\times10^5\). The Bayesian posterior distribution yielded a 95\% credible interval of \([5.63\times10^5,\;7.92\times10^5]\). This estimate is larger than that obtained from the Zero-Inflated Poisson model reported by (\cite{kery2020applied}), whose corresponding $95\%$ interval was (\([4.91\times10^5,\;7.06\times10^5]\)). On the other hand, the bootstrap approximation produced a substantially wider 95\% confidence interval of \([4.28\times10^5,\;9.00\times10^5]\). The lower bound of the bootstrap interval was close to that obtained under the Zero-Inflated Poisson model, despite being derived from the simpler Poisson abundance model, reducing the risk of overly optimistic assessments of population size.

By applying the leave-one-out diagnostic, we identified several sampling sites that had a large influence on the estimated national population size. Left-side of Fig. \ref{fig:LOO} shows the change in the posterior mean of the total population size when each of the 263 sampling sites was removed from the analysis. Most sites had only a negligible impact: removing approximately $80\%$ of the sites changed the posterior mean less than $1\%$. In contrast, five sites altered the posterior mean by approximately $4\sim 6\%$. An important feature of the posterior-covariance diagnostic is that influential observations depend on the target quantity of interest. Right-side of Fig. \ref{fig:LOO} presents the same leave-one-out analysis for another posterior quantity $A(\theta)=\sum_{i=1}^{263}N_i$. The set of influential sites differs from that shown in right-side of Fig. \ref{fig:LOO}, demonstrating that observations influential for one ecological quantity are not necessarily influential for another. 

The posterior-covariance method substantially reduced the computational time required for uncertainty quantification compared with both refitting the Bayesian model using MCMC and performing parametric bootstrap for the corresponding maximum-likelihood model fitted with the \texttt{unmarked} package. A single run of the Bayesian and maximum-likelihood models required approximately 1.2 h and 30 s, respectively. Consequently, 5,000 bootstrap replications would require approximately 250 days for MCMC refitting and 50 h for parametric bootstrap. In contrast, the posterior-covariance method required only about 60 s to generate 5,000 bootstrap replicates for a given quantity $A(\theta)$, most of which was spent calculating the marginal likelihood. Thus, the computation of the posterior-covariance method is approximately 360,000 and 3,000 times faster then MCMC refitting and parametric bootstrap.

\begin{table}
\caption{
Comparison of the computational time for each methods and its cost, relative to the posterior-covariance method.
}
\centering
\begin{tabular}{lcc}
\toprule
Method & Runtime & Relative cost \\
\midrule
MCMC refitting &
$\sim250$ days &
$360{,}000\times$ \\
MLE reffiting &
$\sim50$ h &
$3{,}000\times$ \\
Posterior-Covariance method &
$\sim60$ s &
$1\times$ \\
\bottomrule
\end{tabular}
\end{table}

\section{Implementation and Software}
The introduced methods are implemented in the \texttt{poscosea} package for \textsf{R}, which is publicly available on GitHub:

\begin{verbatim}
install.packages("devtools")
devtools::install_github("OhkuboYusaku/poscosea")
\end{verbatim}

The method assumes two primary inputs obtained from MCMC output:

\begin{itemize}
  \item theta: a vector of posterior draws $\theta = (\theta^{(1)}, \dots, \theta^{(S)})$,
  \item loglik: a log-likelihood matrix $\log p(y_i \mid \theta^{(s)})$, where rows correspond to posterior draws $s=1,\dots,S$ and columns correspond to observations $i=1,\dots,n$.
\end{itemize}

This package does not require MCMC samples to be drawn in a particular package environment. Instead, it only assumes that posterior samples and observation-level log-likelihood values can be extracted from the fitted model. The  methods can be applied to models fitted using \texttt{BUGS}, \texttt{JAGS}, \texttt{Stan}, or any other MCMC software. 

Given posterior samples and log-likelihood evaluations, the approximation is used to compute the posterior mean and observation-wise influence values. The approximation is implemented via

\begin{verbatim}
fit <- ijk(theta, loglik)
\end{verbatim}

The resulting object contains:
\begin{itemize}
  \item $\hat{\theta}$: posterior mean estimate,
  \item $u_i$: influence values for each observation $i$.
\end{itemize}

Leave-one-out (LOO) estimates are obtained as a special case of leave-$k$-out approximation with $k=1$:

\begin{verbatim}
loo <- ijk_lko(
  estimate = fit$estimate,
  influence = fit$influence,
  k = 1
)
\end{verbatim}

More generally, leave-$k$-out estimates can be computed for arbitrary $k$, providing a way to assess sensitivity to subsets of observations. The influence structure can be visualized using diagnostic plots. For the leave-one-out case, each point represents the change in the estimate when a single observation is removed:

\begin{verbatim}
plot(loo)
\end{verbatim}

Large absolute deviations indicate observations with strong influence on the posterior summary. This visualization is particularly useful for identifying leverage points and assessing robustness of the posterior mean estimate.

Then, bootstrap replicates are generated using multinomial reweighting of influence values:

\begin{verbatim}
boot <- ijk_bootstrap(
  estimate = fit$estimate,
  influence = fit$influence,
  B = 5000
)
\end{verbatim}

The resulting distribution is summarized using standard quantities such as bias, standard error, and percentile-based confidence intervals.

\begin{verbatim}
print(boot)

#> 
#> IJK Bootstrap Summary
#> 
#> Original estimate: -0.08 
#> 
#> Bootstrap replicates: 5000 
#> 
#> Bias: 0.038 
#> Bootstrap SE: 1.927 
#> 95% percentile interval:
#> [-3.171, 4.511]
\end{verbatim}

\section{Discussion}
In this article, we introduced computationally efficient approximations to leave-k-out inference and bootstrap resampling for Bayesian models that require MCMC sampling. We evaluated the performance of the posterior-covariance methods through both simulation studies and real-data applications, including hierarchical models commonly used in ecological research.

The primary advantage of the framework is that it avoids re-fitting of models when assessing the sensitivity of posterior inferences to the removal of observations or when approximating bootstrap variability. By exploiting posterior covariance representations of local sensitivity, this approach provides direct approximations to changes in posterior summaries without requiring additional MCMC runs. This reduces the computational burden associated with model assessment and sensitivity analysis. By replacing hundreds or thousands of repeated MCMC runs with covariance-based approximations, the posterior-covariance framework offers a potentially more time and energy efficient alternative for routine sensitivity assessment and uncertainty quantification.

The posterior-covariance method is applicable to a wide range of Bayesian models, provided that the log-likelihood contribution of each observation can be evaluated. As demonstrated in Section 2, the theoretical framework relies on relatively mild assumptions and does not require any specific model formulation or distributional family, such as normal, binomial, or Poisson models. Consequently, the posterior-covariance approach can be readily incorporated into existing Bayesian analyses without modifying the underlying model structure. This generality allows researchers to retain the flexibility of Bayesian modeling while obtaining computationally efficient approximations for sensitivity analysis and uncertainty assessment. 

Despite its broad applicability, the posterior-covariance approach may be particularly useful for N-mixture models. In the standard formulation, latent abundance is assumed to follow a Poisson distribution, while the detection probability is to a binomial distribution. Previous studies, however, have noted that these assumptions can be problematic because ecological count data often exhibit over-dispersion. For example, (\cite{knape2018sensitivity}) investigated the sensitivity of N-mixture models to misspecification and proposed goodness-of-fit diagnostics based on residual analyses. One possible remedy is to replace the probability distribution (e.g., negative-binomial or zero-inflated distribution for abundance model and the beta-binomial distribution for the detection model). Nevertheless, these approaches are not always recommended, as it can lead to unrealistically large abundance estimates or unstable inference (\cite{kery2020applied}). Practitioners often continue to use the Poisson formulation even when some degree of model misspecification is suspected. Parametric bootstrap is commonly used to assess uncertainty in fitted N-mixture models, but this approach generates replicated datasets under the assumption that the fitted model is correct. As a result, uncertainty arising from unmodeled over-dispersion or other forms of model misspecification is not reflected in the resulting inference. When a misspecified model is intentionally employed despite concerns about potential over-dispersion and/or variable detection probability, it is desirable to assess how sensitive posterior inferences are to individual observations and departures from model assumptions. The posterior-covariance method provides a framework for such sensitivity analyses and can contribute to more reliable uncertainty assessment under mild model assumptions.

Several limitations remain to be addressed in future research. Although the posterior-covariance framework is broadly applicable in theory, its practical performance should be investigated for models that are commonly used in ecology and related disciplines. For occupancy models and capture--mark--recapture models, approaches similar to those used for the N-mixture model can be employed. In contrast, applying the posterior-covariance method to state-space models may be substantially more challenging. For some inferential targets, calculation of the observation-level likelihood might be required by integrating over high-dimensional latent state variables. Although some special cases, such as linear Gaussian state-space models, admit efficient algorithms for marginal likelihood evaluation through Kalman filtering, many ecological state-space models are nonlinear and/or non-Gaussian (\cite{auger2021guide}). In such settings, likelihood evaluation may become computationally intensive, potentially limiting the practical advantages of the posterior-covariance approach. Developing efficient approximations for these models represents an important direction for future work.

The approximated bootstrap implemented in poscosea does not provide a panacea; rather, it inherits the limitations of the original bootstrap methods. Theoretically, the bootstrap captures the sampling variability of a statistical procedure without requiring a fully specified parametric model, by assuming that the observed data are random samples from the true data-generating process under several regularity conditions (\cite{efron1994introduction}). Its theoretical foundation is closely related to the Glivenko–Cantelli theorem, which establishes the uniform convergence of the empirical distribution function to the true distribution function. In practice, however, the accuracy of bootstrap methods depends strongly on the sample size because the empirical distribution may not adequately approximate the underlying data-generating distribution in finite samples. When the sample size is small, the approximated bootstrap should be applied with caution.

The approximated bootstrap could be further improved by correcting for bias induced by the prior distribution. An important limitation of bootstrap evaluation of Bayesian estimators is that it does not account for definitional bias (\cite{efron2015frequentist}, \cite{iba2024bias}). While this is not an issue for Fisher-consistent estimators such as the MLE, it can lead to substantial shifts in confidence intervals and reduced coverage accuracy when the prior is informative. Although this issue is not specific to the approximation methods considered in this paper, it may become more relevant in practice because these methods facilitate the bootstrap evaluation of Bayesian estimators. An approach to correcting for definitional bias using posterior covariance is proposed by (\cite{iba2024bias}).

Assessing sensitivity to prior distributions would also be an important topic in Bayesian inference, although this article has focused on sensitivity analysis with respect to the likelihood. Beyond their conventional role in representing prior knowledge or "degrees of belief", prior distributions often serve as essential modeling components for capturing complex dependence structures, such as temporal autocorrelation, spatial correlation, and phylogenetic relatedness (\cite{hadfield2010general}). In these settings, posterior inferences may depend substantially on prior assumptions, and the resulting sensitivity can raise concerns regarding reliability of the conclusions. This issue becomes particularly important in high-dimensional settings, where model complexity often increases with the size of the dataset. In such cases, prior specifications can exert a non-trivial influence on posterior estimates, even when relatively large amounts of data are available. Extending the posterior-covariance framework to quantify sensitivity with respect to prior distributions would therefore be a valuable direction for future research. Such developments could provide practical diagnostics for evaluating the robustness of Bayesian inferences and contribute to more reliable analyses in complex hierarchical models.

\clearpage
\printbibliography
\end{document}